# Business Email Compromise (BEC) and Cyberpsychology

Alessandro Ecclesie Agazzi
*Department of Computing and Informatics*
Bournemouth University
Poole, United Kingdom
alessandro@ecclesieagazzi.com

*Abstract*— This paper gives a brief introduction about what BEC (Business Email Compromise) is and why we should be concerned about. In addition, it presents 2 examples, Ubiquity and Peebles Media Group, which have been chosen to analyse the phenomena of BEC and underpin how universal BEC threat is for all companies.

The psychology behind this scam has been, then, studied. In particular, the Big Five Framework has been analysed to understand how personality traits play an important role in Social Engineering-based attacks. Furthermore, the 6 basic principles of influence, by Cialdini, have been presented to show which strategies are adopted in such scam.

The paper follows with the analysis of the BEC impacts and the incident evaluation and, finally, with the description of some precautions, that companies should undertake in order to mitigate and reduce the likelihood of a Business Email Compromise.

*Keywords*— Business Email Compromise, Social Engineering, Cyber attacks.

## I. INTRODUCTION

BEC (Business Email Compromise) is a sophisticated email fraud which targets companies and businesses which usually work with foreign suppliers and use wire transfers as regular way to transfer funds (Bakarich and Baranek, 2019) (FBI, 2019) (Connell, 2019).

These scams very often include the compromise of legitimate business email accounts to perform unauthorised transfers of funds, although sometimes other variations imply requesting personal information or gift cards (Bakarich and Baranek, 2019) (FBI, 2019) (FBI, 2017).

First labelled an "emerging threat" in 2013 when FBI started tracking cases, BEC is increasingly sophisticated and it constitutes now a global issue (FBI, 2017) (Peter & Associates, 2019) (Barracuda, 2020). According to FBI, "BEC is a severe and global threat, and criminals are ever-more honing their techniques to exploit unsuspecting victims" (FBI, 2017). It's estimated that in 2019, according to a study conducted by Beazley Breach Response Services, BEC has increased by 133% compared to the previous year (Tuttle, 2018) (Barracuda, 2020).

In addition, it also underlined that almost the 25% of all the incidents reported by firms so far are attributable to BEC (Barracuda, 2020).

There are 5 main types of BEC scams (Remorin, Flores and Matsukawa, 2020) (Peter & Associates, 2019); the most common one is the CEO Fraud by which criminals impersonate high-level business executives by either compromising or spoofing their e-mail accounts. The others are "the bogus invoice scheme", "the account compromise scheme", "the attorney impersonalisation scheme", and "the data theft scheme".

BEC exploits 2 types of weaknesses within an organisation (Peter & Associates, 2019):

• Technical vulnerabilities. The exploitation of them is through hacking and it can be mitigated with firewall, servers and anti-intrusion systems (Peter & Associates, 2019).

• People vulnerabilities, "the weakest link". Bad actors are aware that exploiting people, using psychological tricks, is easier than exploiting the system, using only technical skills (Peter & Associates, 2019) (Ferreira et al., 2015).

There are specific tactics, more focused on psychology than computer science, as this paper will show, that criminals employ to exploit the victims (Ferreira et al., 2015). By analysing what about a communication inspires confidence in the recipient, hackers target areas that will make people trust the correspondence. These tactics are known as SE (Social Engineering), and BEC belongs to it (Peter & Associates, 2019).

The first real-world example of BEC incident, this paper is going to deal with, comes from a U.S. public company, Ubiquiti Networks (Bakarich and Baranek, 2019), a technology business which sells products aimed to improve and maintain virtual private networks, surveillance systems and network security (Ubiquity, 2018).

In addition, it will be provided another case where the victim of this attacks has been Peebles Media Group "one of Scotland's leading independent media companies with a diverse portfolio of high profile B2B and consumer titles, as well as an events division that holds several key industry awards and trade events" (Peebles Media Group, 2017).

These cases have been selected as relevant to fully understand the psychology and the impact behind a BEC attack; These cases emphasise how universal BEC threat is for all companies (Connell, 2019); in spite Ubiquiti's focus on Internet and Host security, it fell victim to one of the most common kinds of BEC attack (Bakarich and Baranek, 2019).

Finally, the second example shows how the impacts could be even more severe than the losses (Wanca and Cannon, 2020) (Sussman, 2019).

The Pebbles Media Group's employee, who had been cheated by a CEO fraud, was fired and even been sued for more than 130.000 £ by her company to recover the money lost in this cyber attack (Sussman, 2019) (BBC News, 2019) (Moskvitch, 2019).

## II. INCIDENT IDENTIFICATION

**UBIQUITY CASE**

Ubiquity is a company which sells products to ensure online/offline protection and security and it has wholly owned subsidiaries in Hong Kong, China, Lithuania, Poland and a few other countries around the world (Ubiquity, 2018).

On the 19th of May 2015, a person, working in the finance department of Ubiquity, received an email which, he thought, came from an executive of the same company. (Bakarich and Baranek, 2019) (Forbes, 2016).

Said email instructed the employee to transfer money due to the acquisition Ubiquity was conducting at that period. In addition, it also stated that on the requested transfer, Tom Evans, an outside attorney from the international law firm of Latham & Watkins, would follow up with the instructions for approving the payments. The finance employee had to follow Tom's instructions (Bakarich and Baranek, 2019).

Following the first email, a second one was received by the employee on the same day and it instructed him to proceed immediately with the payment (Bakarich and Baranek, 2019). The victim, as per the instruction, wired the money from Ubiquiti's Hong Kong bank account.

Over the next 17 days, the victim followed the instructions of the fake executive and made 14 transfers ($46.7 million) to foreign accounts which were held by third parties in China, Russia, Poland and Hungary (Ubiquiti, 2015) (Forbes, 2016).

However, at that time, Ubiquiti was not conducting any acquisition. Investigators found out later that those emails, instructing the victim to wire transfer money, were sent by criminals (Bakarich and Baranek, 2019).

They looked similar to a real one and included the electronic signature of the company Latham & Watkins', but they had been sent from an e-mail account which ended with the domain: @consultant.com (Ubiquiti, 2015) (Forbes, 2016).

On the 6th of August 2015, the company Ubiquity informed the Security and Exchange Commission (SEC) that they suffered a BEC scam, involving "employee impersonation and fraudulent requests from an outside entity" (Ubiquiti, 2015) (Bakarich and Baranek, 2019) (Forbes, 2016).

**PEEBLES MEDIA GROUP**

Peebles Media Group is a leading independent publisher with experience in the Scottish B2B and B2C markets. They offer media solutions across digital platforms, print and events. B2B titles include Scottish Licensed Trade News, Scottish Grocer, Project Scotland, Project Plant, Packaging Scotland and Envirotec. B2C titles include Homes & Interiors Scotland and Tie the Knot Scotland (Peebles Media Group, 2017).

The victim who was cheated by BEC scam was Reilly, a finance department employee. She got different emails which seemed to come from the company director Yvonne Bremner, her manager, in the Summer of 2015 while he was on vacation in Tenerife (Sussman, 2019) (BBC News, 2019) (Moskvitch, 2019).

Those emails asked for the help of Reilly to transfer $200,000 via wire transfer, from one account to another, as he was temporary unable because on holiday (Sussman, 2019).

Reilly made the transfer as requested, unaware that those emails, in fact, were sent by criminals who knew Bremner was off (BBC News, 2019).

Some days later, another finance employee contacted Yvonne Bremner in Tenerife and discovered he had not requested that transfer (Sussman, 2019).

They, thus, found out the company had been victim of the BEC cyberattack (Moskvitch, 2019).

Just after the discovery, the bank of the company managed to recover a part of the money by blocking the transaction. However, criminals succeeded in keeping $138,000 of the transfer (Sussman, 2019).

Peebles Media Group responded to the incident by firing Reilly and then suing her for the lost $138,000 (Sussman, 2019) (BBC News, 2019).

## III. INCIDENT ANALYSIS

Every BEC attack, like for those occurred in the already presented examples, follows a similar pattern and rules aimed to increase the likelihood and the success of it (Barracuda, 2020).

**PSYCHOLOGY BEHIND THESE ATTACKS**

BEC attack aims to overcome the security controls by exploiting weaknesses in human behaviour and decision making (Uebelacker, 2020). These scams usually rely on a mixture of tactics to influence and persuade decision making such as authority, time pressure, and tone (Dhamija, Tygar and Hearst, 2006). BEC, along with phishing, claims to come from a reliable and trustful source, with corporate logo and with its name which seems legitimate and trustworthy (Public-Private Analytic Exchange Programme, 2018) (Wanca and Cannon, 2020) (Dhamija, Tygar and Hearst, 2006). In addition, BEC email often contains an element of time pressure and urgency and it may also prey on user's fear of something. The tone of these messages usually involves persuasive and polite statements to further manipulate the decision making (Public-Private Analytic Exchange Programme, 2018) (Segura, 2017).

This section is going to examine the correlation between the response to BEC scams and the Big five personality traits (Halevi, Lewis and Memon, 2013).

In the classical decisional theory, decision making under risk is assumedly based on logic. As a result, reasonable people should take rational choices. However, it has been showed that humans' decision tends to be biased (Uebelacker, 2020).

A criminal may, hence, manipulate the decision-making process (Guadagno and Cialdini, 2005) (Cialdini, 2009) (Sagarin et al., 2002).

BEC fraud appeals to different human vulnerabilities, like the desire for immediate gain, to help others and to be thanked by the manager. It has been shown that certain people have "victim personalities" which lead them to be more vulnerable to the Social Engineering scams (Halevi, Lewis and Memon, 2013).

**BIG FIVE FRAMEWORK**

In psychology, personality is defined as a person's relatively stable feelings, thoughts, and behavioural patterns. These are predominantly determined by inheritance, social and environmental influence, and experience, and are therefore unique for every individual (McCrae and John, 1992).

Personality is a consistent pattern of how people respond to stimuli in their environment and their attitude towards different events. The Five Factor model is used for multidimensional measures of personality: Neuroticism, Extroversion, Openness, Agreeableness, and Conscientiousness (Uebelacker, 2020).

The literature in this field provides a relation between personality traits and SE in general. There is a correlation between personality traits and the probability of being cheated by a social engineering attack. For some traits of personality, indeed, the likelihood of the successful attack may increase or decrease. That's why BEC attack, which is a social engineering attack, it's a high targeted attack where victims are studied carefully (Barracuda, 2020).

Here the big five framework traits (Uebelacker, 2020) (Halevi, Lewis and Memon, 2013):

- Conscientiousness: since conscientious people adhere to existing rules, it is assumed that they are more vulnerable to Social Engineering techniques that exploit rules, social norms, and policies (Halevi, Lewis and Memom, 2013).

- Extraversion: SE attacks using liking or social proof can work well on extraverted individuals as they rely on social aspects because extraversion relates to sociability, a sub-trait of extraversion (Uebelacker, 2020). The excitement seeking is one sub-trait that can lead to greater vulnerability for the scarcity principle – getting something scarce is usually described as exciting (Uebelacker, 2020).

- Agreeableness: individuals who are more trusting raise fewer concerns about privacy invasion by location-based services, which we assume to be generalisable to fewer privacy concerns. We predict higher SE vulnerability because of the higher likelihood of disclosing private information if a social engineer established a trust relationship (Uebelacker, 2020).

- Openness to Experience: openness to experiences and strong fantasy lead to higher susceptibility. On the other hand, openness is associated with technological experience and computer proficiency. Therefore, openness reduces SE vulnerability as more digitally literate users better detect SE attacks (Halevi, Lewis and Memom, 2013) (Uebelacker, 2020).

- Neuroticism: Parrish et al. (2019) propose that computer anxiety, which is associated with neuroticism, may protect regarding to computer-based SE attacks like phishing because neurotic users act with more caution. In general, they established that neurotic individuals are less susceptible to most SE attacks. Neuroticism can act as a barrier since the underlying pessimism often assumes the worst in any situation (Uebelacker, 2020).

**INFLUENCE**

Personality traits play an important role when it comes to social engineering; however, there are some tricks used to manipulate and influence the victims (Uebelacker, 2020) (Ferreira et al., 2015). Some strategies of Social Engineering, adopted in areas such as marketing, were analysed by Cialdini. He grouped them into 6 categories, called "the 6 basic principles of influence" (Ferreira et al., 2015).

To demonstrate that Cialdini's principles are applicable in context of BEC and more generally to SE, they have been applied to Ubiquity and Peebles Media Group's cases (Uebelacker, 2020).

The six principles consist of (Uebelacker, 2020):

- Authority: People tend to comply to authorities even if they are persuaded to behave unethically (Mosley, 2020) (Jakobsson, 2014) (Uebelacker, 2020). There are 2 kinds of Authority:
    o Authority based on expertise, such as police.
    o Authority based on the position occupied within a company, like the finance executive for the Ubiquity's case or the manager for the Peebles Media Group's one.

- Social Proof: Adapting behaviours or beliefs to be "socially accepted"; it implies a high level of trust towards people who share similar ideas (Uebelacker, 2020).

- Scarcity: People tend to assign more value to less available opportunities. This is applicable also for requests which claim to have a high sense of urgency, where victims must act quickly. This principle is correlated also to the idea of fear ("I act quickly because, otherwise, I get in trouble") and to the one of distraction (acting quickly could lead to mistakes or could make victims overlook the scam cues) (Brehm, 1966).

- Commitment & Consistency: People strive for consistency in their commitments. In addition, getting customers/colleagues to publicly commit makes it more likely they will follow through (Guadagno and Cialdini, 2005).

- Liking: People prefer to say 'yes' to what they know and like. The employees tend to accept the request coming (or pretending to come) from their managers (Bujold, 2002).

- Reciprocity: Reciprocity makes establishing trust with others and refers to our need for equity. The power of reciprocity can be so high that the target would return an even greater favour than what was received (Public-Private Analytic Exchange Programme, 2018).

The following table shows how these principles of persuasion have been applied to the cited cases:

|  | Ubiquity's case | Peebles Media Group's case |
|---|---|---|
| **Authority** | The emails received by the employee came from an executive with electronic signature and Latham & Watkins' details. | The email received by the employee came from her boss and had company logo, the mail has been spoofed and appeared to be very similar to the real one. |
| **Social Proof** | The task of the victim implied the collaboration with an outside attorney, Tom Evans from the international law firm of Latham & Watkins. | The employee complied with the task because "other people would have done it" |
| **Scarcity** | The victim received, on the same day of the first email, a second one which instructed him to make the first payment immediately. | The victim received an email from "her boss" which ordered her to make the payment immediately. |
| **Commitment & Consistency** | The employee wanted to act for the wealth of the company and was committed to his position. | The employee wanted to act for the wealth of the company and was committed to her position. |
| **Liking** | The employee wanted to help the company and acting quickly would have been appreciated by them. | Reilly wanted to act quickly and help his manager who was currently on holiday. |
| **Reciprocity** | The employee got involved in an important task and wanted to reward the company by acting as per instructions | Reilly got involved in a task from her boss and wanted to reply straight |

## IV. FURTHER INCIDENT ANALYSIS

The Ubiquity and Peebles Media Group's examples showed how the impact could be potentially violent when it comes to falling under a BEC scam. In addition, the data obtained through these scams can also lead to further victimisation (Wanca and Cannon, 2020). For instance, if an employee's personal information (such as name, email address, phone number) is posted online, other criminals could utilise this information to commit other crimes against the victim (e.g., stalking, harassment, burglary).

Based on the analysed cases and the online literature regarding BEC attacks, here below the most relevant impacts this scam leads to (Peter & Associates, 2019) (Wanca and Cannon, 2020):

- Lost Time: recovering from a BEC attack can be confusing, time confusing, and generally inconvenient for victims. Depending on the type of damage caused by it, victims can spend anywhere from a few hours to many months or year resolving the associated problems.
- Trauma: BEC, along with SE attacks, can cause significant emotional distress (e.g., denial, loss of trust, frustration, fear, anger, powerlessness, helplessness, embarrassment, depression, sleep disturbances). Some theorise that cybercrime victimisation, such as identity theft, can be more harmful to victims than crimes like property theft; one can replace property, but it is not possible to acquire a new identity. Further, BEC victims can experience secondary victimisation by others who blame the victims for falling for the attack.
- Financial loss: victims can incur both direct (i.e., value of goods, services, or cash obtained) and indirect (e.g., legal fees, bounced checks, postage) financial loss resulting from BEC attacks. In the Ubiquity case, the financial loss was $46.7 million. In addition, the employers of BEC victims can experience financial losses related to decreased productivity (see below), business disruption, isolating malware and credential compromises, and the cost of data breaches.
- Social Consequences: victimisation can cause strain on personal and family relationship and reputational damage. For example, if cybercriminals gain access to a victim's email, they can uncover information about personal relationship or embarrassing photos or videos that may be leaked to the public. Family or friends could also become the targets of cybercriminals.
- Business consequences: both intellectual property and customer data can be at risk when a BEC attack occurs. In addition to financial loss, such attack can damage the reputation and credibility of a business. Consumer may loose trust in the business, which can lead the company to loose its customer base.

- Lost productivity: the time it takes to recover from a BEC scam and the trauma inflicted can result in decreased employee productivity. It is estimated that non-IT employees spend an evarage of 4.16 hours per year dealing with such attacks.

## V. INCIDENT EVALUATION AND SUMMARY RECOMMENDATIONS FOR RISK MITIGATION

Ubiquity and Peebles media group's cases emphasise how universal BEC threat is for all companies. Both BEC attacks were performed using the 6 Cialdini's persuasion principles and addressing specific employees whose personality traits would have facilitated such outcome.

Getting back to section 3, some personality traits, such as neuroticism, Conscientiousness, Extraversion and Agreeableness have been found to be correlated to BEC (Parrish et al., 2009).

Easy-going, careless, outgoing, nervous, naïve people are found to be, indeed, the most hit targets by criminals and the cited attacks were successful due to particular victim's personality traits, in particular (Parrish et al., 2009):

- Extraversion and Conscientiousness are correlated to BEC vulnerability and, according to the study of (Halevi, Lewis and Memon, 2013), this is also gender-based, being particularly significant for women rather than men.

- Openness to experience, from one hand, may lead to a higher tendency of sharing and disclosing personal valuable information; however, from the other hand, it could also lead to an increase of the experience in IT, reducing thus the likelihood of falling victim to BEC scheme (Halevi, Lewis and Memon, 2013).

- Agreeableness may also constitute a vulnerability because friendly and compassionate people could perform a request to conform with the company and/or manager (Halevi, Lewis and Memon, 2013).

In order to mitigate and minimise the outcome of such scams, in Ubiquity and Peebles Media Group's cases, some steps should have been taken (Peter & Associates, 2019):

- Check and Communication: Both employees should have checked the unexpected emails sent by high-level executives and got a secondary verification of the request asking directly to the person in charge of it (Remorin, Flores and Matsukawa, 2020), "The best way to avoid being exploited is to verify the authenticity of requests to send money by walking into the CEO's office or speaking to him or her directly on the phone. Don't rely on e-mail alone." (FBI, 2017). They also should have checked the changes in vendor payment details by using a secondary sign-off.

- Employee awareness (Remorin, Flores and Matsukawa, 2020) (Biskup and Weil, 2020). Both Ubiquity and Peebles Media Group should have raised an awareness training aimed to understand and detect these scams.

-

## VI. RECOMMENDATIONS

The previous section dealt with how said scams may have been mitigated; this one, instead, is going to present some recommendations and tactics to reduce and minimise BEC attack, a very common threat for companies and enterprises (Peter & Associates, 2019) (KnowBe4, 2020).

As a result, Deloitte performed a research on BEC attacks and focused on the mitigation of such attacks:

In particular, they stressed different steps to undertake in companies in order to mitigate and reduce the likelihood of a Business Email Compromise (Remorin, Flores and Matsukawa, 2020) (Biskup and Weil, 2020):

1. Train of employees in order to detect BEC attacks.

    Employees have to be trained about how BEC scheme works and how criminals structure their attacks using real and harmful persuasive techniques (Remorin, Flores and Matsukawa, 2020).

    Especially important in preventing BEC is employees training which makes them be aware about risks, implications and tactics used by criminals.

    Moreover, a well-structured training course fosters a sense of responsibility in the organisation (Biskup and Weil, 2020).

2. Creation of a culture of compliance.

    BEC scams may not be mitigated just by training, criminals keep improving their tactics making the identification of them very difficult. As a result, compliance should go in parallel with training (Remorin, Flores and Matsukawa, 2020).

    Compliance would allow the employees to follow up a secure procedure, which in some cases would unmask attempted frauds, although the request comes from the chief officer of the company (Biskup and Weil, 2020).

3. Build a technical defence.

    Although BEC is effective due to its sophisticated psychological pattern, it is not considered technically advanced; its origin, indeed, usually comes from phishing or spoofing email (Public-Private Analytic Exchange Programme, 2018) (Biskup and Weil, 2020).

    These scams could be prevented by implementing application-based multi-factor authentication (MFA) and virtual private networks (VPNs).

    In addition, the communication via e-mails could be secured by encrypting the emails exchange. This would allow a protection against spoofing and phishing emails (Biskup and Weil, 2020).

## VII. CONCLUSION

This paper gives a brief introduction about what BEC is and why we should be concerned about. In addition, the first section introduces the 2 examples which have been chosen to analyse the phenomena of BEC: Ubiquity and Peebles Media Group's cases in order to underpin how universal BEC threat is for all companies.

In the second section, instead, the psychology behind this scam has been studied. In particular, the Big Five Framework has been analysed to understand how personality traits play an important role in social engineering-based attacks. In addition, the 6 basic principles of influence, by Cialdini, have been presented to show which strategies are adopted in such scam.

The third chapter aims to further analyse the incidents by dealing with the impacts that BEC scam may have on enterprises, businesses and, especially, on people. Among loss of time, money, productivity, it could lead to emotional distress, frustration, anger, embarrassment or victimisation.

The incident evaluation section aims to describe the personality traits of the victims of such cases. Ubiquity and Peebles Media Group's victims are, in fact, unknown but, basing on the literature of cyberpsychology, they have been analysed and the most hit target's traits have been presented. Moreover, some precautions which should have been taken are stressed.

Finally, the last section presents 3 steps that companies should undertake in order to mitigate and reduce the likelihood of a Business Email Compromise.

## VIII. LIMITATIONS AND FUTURE PROJECTS

This paper presents some limitations; the principal one regards the documentation of the occurred incidents. The volume of literature regarding the Ubiquity and Peebles Media Group's scams is very low and it lacks important details, such as the description of the victims' personality, which could have been very useful for the psychological analysis of said cases.

In addition, scammers are not presented thus it has been impossible to introduce an important field of cyberpsychology: the bad actor, his psychology and why he behaves in such way.

While doing researches for this paper, other interesting topics have been raised; however, due to the words limit, it has not been possible to analyse another related domain of this topic. Future projects relating BEC will focus on:

- Comparison of the personality traits with the six principle of persuasion and analysis of how every trait responds to each principle.
- Analysis of the scammers and their psychology.
- Analysis of the steps by which the bad actor carries out the attack, from the reconnaissance to the exploitation (Public-Private Analytic Exchange Programme, 2018).
- Comparison of the trends and impacts of BEC before and during the COVID-19 period to understand how people react to such scam during a period of crisis.